# Motion of a deformable drop of magnetic fluid on a solid surface in a rotating magnetic field

A. Zakinyan, O. Nechaeva, Yu. Dikansky

Department of General Physics, Stavropol State University, 1 Pushkin St, 355009 Stavropol, Russia

Address correspondence to A. Zakinyan:
phone: +7-8652-57-00-33,
postal address is presented above,
e-mail: zakinyan.a.r@mail.ru

**Abstract.** The behavior of a magnetic fluid drop lying on a solid horizontal surface and surrounded by a nonmagnetic liquid under the action of a uniform magnetic field which is rotating in a vertical plane with low frequency (of the order of 1 Hz) has been investigated experimentally. Shape deformation and translatory motion of the drop were observed and studied. The drop translation velocity for different field amplitudes and field frequencies has been measured.

## 1. Introduction

The rotational dynamics of liquid drops have been under investigation for more than a hundred years. Experimental investigations in this field were held back for a long time by the difficulty of producing rotating drops of liquid. Use of magnetic fluids offers unique opportunities in investigations of such kind. In this case, the possibility of controlling the behavior of a magnetic fluid drop by external magnetic fields including setting it in rotation and shaping it into various configurations is taking place. And new features of rotational dynamics of liquid drops appear in behavior of drops of a magnetic fluid under the action of external magnetic field.

A magnetic fluid is a colloidal suspension of ultra-fine ferro- or ferri-magnetic nanoparticles suspended in a carrier fluid. Magnetic fluid drops can be set in rotation by applying a rotating magnetic field. The behavior of a magnetic fluid drop freely suspended in a surrounding nonmagnetic liquid and exposed to a rotating magnetic field was studied experimentally and theoretically in many works, e.g. [1–7]. Investigation of the influence of bounding surfaces on rotational dynamics of a drop is of considerable interest. Such investigations for a drop of a magnetic fluid are presented in works [8, 9], where the behavior of the drops lying on a solid or liquid horizontal surface and surrounded by air has been studied. It was observed in [8, 9] that the action of a high-frequency uniform magnetic field which is rotating in a vertical plane leads to a translatory motion of the drops in the direction the field is rolling. The drop shape is slightly strained under the action of a field in this case. The reason of such motion of a drop is that the application of a rotating magnetic field induces a torque on the suspended magnetic nanoparticles in a magnetic fluid. Due to the viscous coupling of the particles to its surrounding carrier liquid angular momentum is transferred to the whole drop.

In contrast to the above mentioned works, in the present work we report for the first time on the experimental study of the behavior of magnetic fluid drop lying on a solid surface and surrounded by nonmagnetic liquid under the action of a low-frequency magnetic field which is rotating in a vertical plane. In our case, the internal rotations of magnetic particles in a magnetic fluid drop are inessential owing to quite slow rotation of a magnetic field. As it was shown in [8], the internal rotations become significant in a high-frequency rotating magnetic field when the field rotation period is close to the characteristic time of Brownian rotational motion of a particle in a viscous medium. Another difference of the case we studied is that the drops shape is strained much more strongly under the action of a magnetic field. Apart from independent interest, the results of such studies can be useful in analysis and modeling of fluids containing deformable particles (polymer solutions, emulsions, blood, etc.).

## 2. Experiments

The experimental setup is shown schematically in Fig. 1. A magnetic fluid drop was placed using a syringe on the flat bottom of the rectangular glass reservoir filled with a glycerin/water mixture. The glycerin/water mixture wets the glass surface better than the magnetic fluid, creating droplets with a large contact angle that appear like beads. The contact angle of the magnetic fluid droplet was measured as ~130°. The diameter of droplet cross-section was 3 mm. Glycerin/water mixture was poured into a glass reservoir up to a height of 2 cm in order to minimize the influence of the upper surface of the mixture on the studied processes. The reservoir with magnetic fluid drop inside was placed in a uniform magnetic field rotating in a vertical plane.

The rotating uniform magnetic field was produced by two perpendicular pairs of Helmholtz coils supplied with alternating current from a two-channel function generator with a phase shift of $\pi/2$. The current amplitude in each pair of coils was controlled independently by two power amplifiers. An oscilloscope served for the monitoring of the parameters of both currents. The magnetic field strength in experiments varied from 0 to 6 kA/m, and the rotation frequency of the field varied from 1 to 4 Hz. The behavior of a magnetic fluid drop was observed by a digital video camera.

In our experiments, we used a kerosene-based magnetic fluid with dispersed magnetite nanoparticles of about 10 nm diameter stabilized with oleic acid. The properties of the magnetic fluid used in experiments are: density is 1400 kg m$^{-3}$, dynamic viscosity is $\eta_i \sim 20$ mPa s, initial magnetic susceptibility ~ 4. Since the size of magnetic nanoparticles suspended in the magnetic fluid is much smaller than the size of a magnetic fluid drop, in this case the magnetic fluid can be considered as a continuous magnetizing liquid medium. The mixture of water and glycerin comprise both components in an equal proportion. The density of the mixture was 1150 kg m$^{-3}$ and its dynamic viscosity was $\eta_e \sim 0.5$ Pa s. The interfacial tension at the boundary between the magnetic fluid and the glycerin/water mixture was $\sigma \sim 20$ mN/m.

## 3. Results

It was observed that the drop shape deformed significantly under the applied magnetic fields used in the current experiments, elongating the drops in the direction of the applied field. It is well known that in a static or low-frequency rotating magnetic field, the magnetic fluid drop which freely floats takes the shape of an ellipsoid of revolution elongated along the field direction. In our case the drop shape is more complicated because of the distortion effect from the solid surface. As an example, a series of snapshots of the drop made in a vertical plane are shown in Fig. 2.

A new interesting phenomenon was observed in the behavior of the magnetic fluid drop. In addition to the shape deformation under the action of a rotating magnetic field the drop moves along the surface of a bottom. The direction of the droplet motion is opposite to the direction the field is rolling. So, if the field rotates anticlockwise, then the drop moves to the right (Fig. 2). The drop moves in a straight line with a constant translation velocity with respect to the solid surface. Under the given experimental conditions we can achieve droplet velocities up to a few millimeters per second. The movie contains a complete experimental observation of the behavior of a magnetic fluid drop.

We have measured the drop velocity for different field strengths and field frequencies. The velocity of the drop was determined by extracting the time a drop takes to travel a distance of 1 cm. To this end, the motion of the magnetic fluid drop was observed from above by means of a video camera. The scale at the bottom of the reservoir (Fig. 1) was used to measure the droplet path length. Fig. 3 shows the obtained experimental dependences of the drop velocity on the magnetic field strength at two different values of the rotation frequency of the field. As is seen, the drop velocity grows with the magnetic field strength increasing.

The frequency dependence of the drop velocity, measured at three different values of a magnetic field strength, is presented in Fig. 4. As is seen, these dependences have maxima. In our experiments the drop reached the maximum velocity at a rotation frequency of about 2 Hz. It was observed in the experiments that the degree of deformation of the droplet diminishes with an increase of the rotating field frequency. And at comparatively high frequencies the drop is deformed slightly and the deformation direction do not follow the field direction. At such frequencies the translatory motion of a drop is not observed. So, it can be concluded that for the observation of a droplet motion, a considerable strain of the droplet is needed. Such situation is taking place if the field frequency is less than the inverse relaxation time of the drop shape. It can explain the decrease of the drop velocity at high field frequencies and appearance of maxima of the dependences shown in Fig. 4.

## 4. Discussion

As it was already mentioned above, a high-frequency rotating external magnetic field induces a rotation of the magnetic nanoparticles in magnetic fluid. This rotation is transferred to the whole drop of magnetic fluid and leads to the translatory motion of a drop if it is placed on a flat surface [8, 9]. The drop motion observed in our experiments presents a new result and quite differs from that investigated in works [8, 9] in two points fundamentally. Firstly, the motion direction we observed is opposite to that which can occur owing to the internal rotations. Secondly, the drop shape is not so important for its motion driven by internal rotations, and the drop maintains a stationary slightly strained shape in that case. On the contrary, in the case studied here, a significant shape deformation and rotation under a low-frequency magnetic field is a necessary condition for the drop motion, as it was shown in the previous section. To confirm this conclusion, let us estimate the droplet oscillations eigenfrequency. According to the small deformation theory, the eigenfrequency can be written as $f^* = 40\sigma(\eta_e + \eta_i)/[2\pi R(16\eta_e + 19\eta_i)(3\eta_e + 2\eta_i)]$, where $R$ is the drop radius. Substituting all parameters, we obtain the droplet oscillations eigenfrequency value equal to $\approx 3$ Hz. This value of $f^*$ is near the magnetic field frequency at which the drop reaches the maximum velocity (Fig. 4). At higher frequencies the droplet shape can no longer follow the magnetic field rotation and the velocity of the droplet translatory motion diminishes.

It is essential to mention that a lot of studies have been made on the subject of motion of a rotating rigid ellipsoid in a liquid medium near a wall (see [10] for a review). In the case of closely-spaced ellipsoidal particle and wall, the particle is acting as a "paddle" and rowing itself across the wall and the resulting motion of the particle is in the direction the particle is rolling (opposite to that observed in our experiments). It should be also noted that in some works (e.g. [3]) the effect of the breakup of a magnetic fluid drop in a rotating magnetic field was reported. In our experimental conditions such effect was not observed.

Recently, some investigations have been undertaken focusing on the manipulation and transport of magnetic fluid droplets for microfluidic applications [11, 12]. For these purposes, magnetic fluid droplets are commonly manipulated utilizing local field gradients, which are locally created by planar coils. We propose that the results of this study can be used for development of a new promising technique of the controlled transport of small amounts of liquid to any desired position on a solid surface. Our driving technique yields a constant droplet velocity on a complete surface and can gain some useful applications in droplet-based microfluidics.

**5. Conclusion**

The main achievement of this paper is that rotating fields can transport magnetic fluid drops. In particular, we have studied a magnetic fluid drop lying on a horizontal glass surface and surrounded by a nonmagnetic liquid under the action of a uniform magnetic field which is rotating in a vertical plane with low frequency. A complex deformation of the shape of the drop accompanied by its translatory motion along the surface was observed for the first time. The drop translation velocity for different field amplitudes and field frequencies has been measured.

**Acknowledgements**

This work was supported by Russian Foundation for Basic Research (project No. 10-02-90019-Bel_a) and by the Federal Education Agency of the Russian Federation in Scientific Program "Development of Scientific Potential of Higher School".

**Figure captions**

Fig. 1. Sketch of the experimental setup: *1* – magnetic fluid drop; *2* – glass reservoir; *3* – glycerin/water mixture; *4* – scale; *5* – Helmholtz coils; *6* – function generator; *7* – power amplifiers; *8* – oscilloscope; *9* – digital video camera; *10* – computer.

Fig. 2. Consecutive snapshots at 0.1 s intervals showing a magnetic fluid drop on a glass surface under the action of a rotating magnetic field. The magnetic field strength is 5 kA/m and the rotation frequency of the field is 1 Hz. For the complete observation of the drop behavior see movie.

Fig. 3. Dependence of the drop translation velocity on the magnetic field strength for two different values of the rotation frequency.

Fig. 4. Dependence of the drop translation velocity on the rotation frequency of the field for three different values of the magnetic field strength.

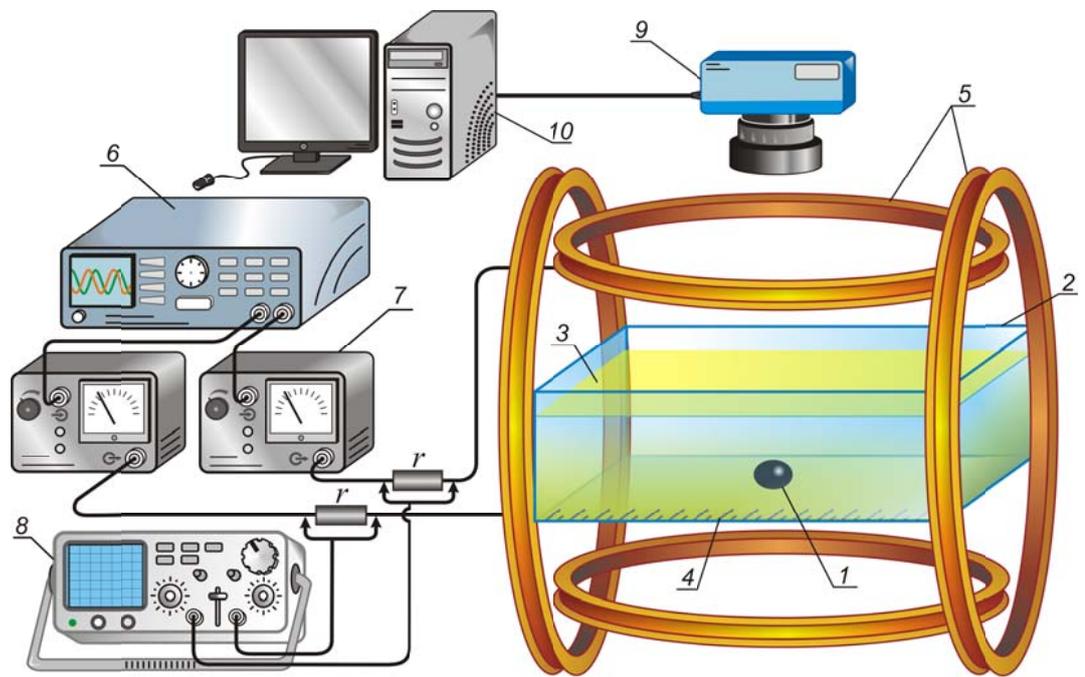

FIG. 1.

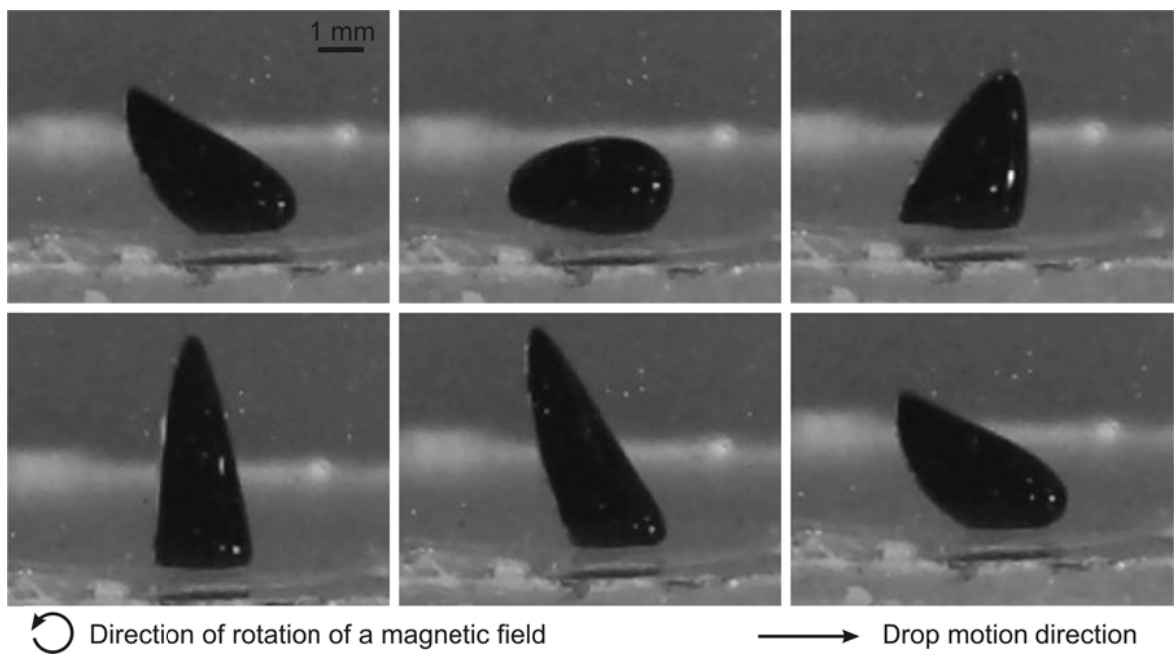

FIG. 2.

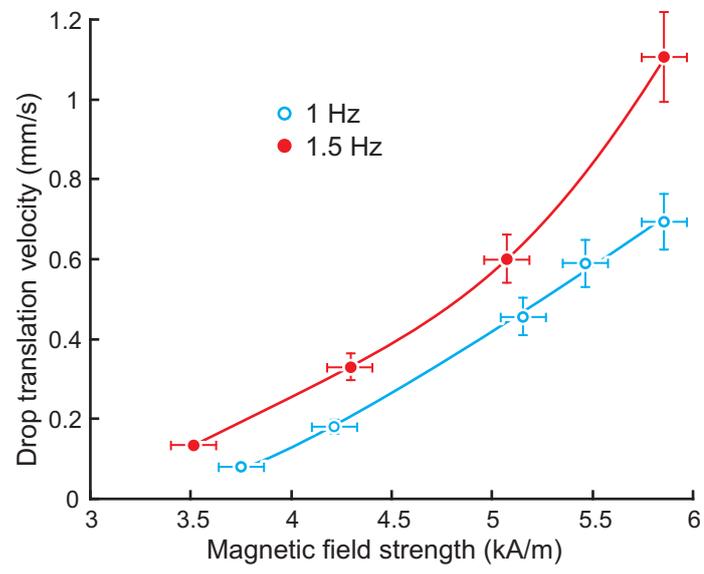

FIG. 3.

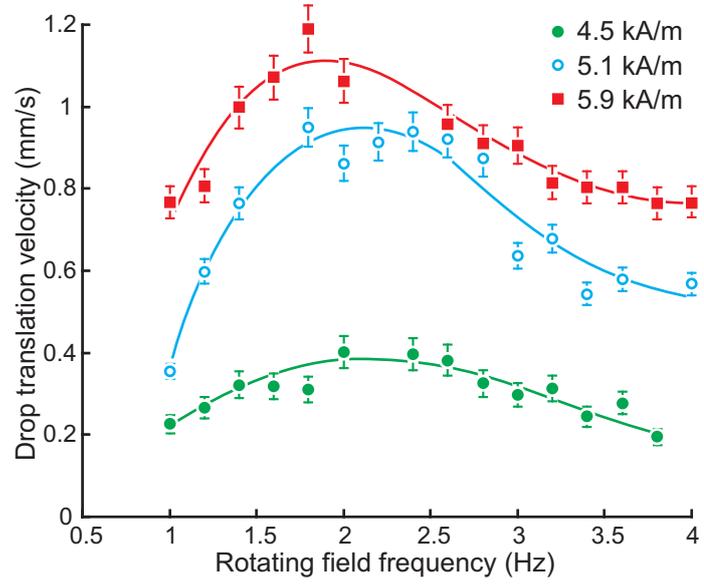

FIG. 4.